\documentclass[10pt,journal]{IEEEtran}
\usepackage{ifpdf}
\usepackage[noadjust]{cite}
\usepackage[hyphens]{url}
\usepackage{graphicx}
\usepackage[cmex10]{amsmath}
\usepackage{mathtools}%
\usepackage{algorithmic}
\usepackage{array}
\usepackage{stfloats}

\begin{document}

\title{A Subject-Specific Four-Degree-of-Freedom Foot Interface to Control a Robot Arm}

\author{Yanpei~Huang,
        Etienne~Burdet*, 
        Lin~Cao*, 
        Phuoc~Thien~Phan,
        Anthony~Meng~Huat~Tiong,   
        and~Soo~Jay~Phee 
\thanks{All authors are or were with the School of Mechanical and Areospace Engineering, Nanyang Technological University, Singapore. Etienne Burdet is also with the Department of Bioengineering, Imperial College of Science Technology and Medicine, London, UK. (*Corresponding authors' emails: e.burdet@imperial.ac.uk;lin.cao@ntu.edu.sg)}}

\maketitle
\begin{abstract}
In robotic surgery, the surgeon controls robotic instruments using dedicated interfaces. One critical limitation of current interfaces is that they are designed to be operated by only the hands. This means that the surgeon can only control at most two robotic instruments at one time while many interventions require three instruments. This paper introduces a novel four-degree-of-freedom foot-machine interface which allows the surgeon to control a third robotic instrument using the foot, giving the surgeon a ``third hand". This interface is essentially a parallel-serial hybrid mechanism with springs and force sensors. Unlike existing switch-based interfaces that can only un-intuitively generate motion in discrete directions, this interface allows intuitive control of a slave robotic arm in continuous directions and speeds, naturally matching the foot movements with dynamic force \& position feedbacks. An experiment with ten naive subjects was conducted to test the system. In view of the significant variance of motion patterns between subjects, a subject-specific mapping from foot movements to command outputs was developed using Independent Component Analysis (ICA). Results showed that the ICA method could accurately identify subjects' foot motion patterns and significantly improve the prediction accuracy of motion directions from 68\% to 88\% as compared with the forward kinematics-based approach. This foot-machine interface can be applied for the teleoperation of industrial/surgical robots independently or in coordination with hands in the future.
\end{abstract}


\section{Introduction}
\IEEEPARstart{H}{uman-machine} interfaces (HMI) play an important role in teleoperation systems. A well-built HMI can accurately identify the commands from the user and provides the user with haptic feedback from the remote environment. Many HMIs for hands are available in the market, e.g., PHANToM (3D Systems, USA), Delta/ Omega/ Sigma series (Force Dimension, Switzerland), etc. These HMIs are under the direct control of the user's hands. They can act as master devices telemanipulating remote robotic arms through the sub-robotic system.

HMIs have been widely used in robotic surgical systems. For example, the well-known Da Vinci system \cite{DaVinci2004} has a built-in master console through which the surgeon tele-manipulates or switches instruments. The RAVEN \cite{Raven2} and MiroSurge \cite{Tobergte2011} robotic systems offer teleoperation and  haptic feedback through commercial hand interfaces PHANToM Omni and Sigma 7, respectively. The flexible endoscopic robotic MASTER \cite{EndoMaster2017} system has similar teleoperation mechanisms for the manipulation of two flexible robotic instruments in a flexible endoscope. While these HMIs provide good manipulation control, they are designed and used for the hands so that the surgeon can only control at most two instruments at one time.

These HMIs are too limited for many surgical operations, which require using three instruments simultaneously. In this case a human assistant is needed to help controlling the additional instrument(s) in coordination with the surgeon. For instance, a camera assistant is needed to adjust the camera field of view during laparoscopic surgery; in a robotic flexible endoscopic surgery, the robotic arms controlled by the surgeon can only work in a small workspace and thus an endoscopist is often needed to move the endoscope which is carrying the robotic arms to adjust the working area \cite{EndoMaster2017}. However, it was observed that surgical performance and efficiency are often limited by the communication delays and errors between the surgeon and the assistant \cite{Nurok2011}, and any mistakes may affect the patient's health. 

A solution to address this issue is let the surgeon control the additional instrument(s) \cite{Multimodal2014}. Our vision is that the surgeon should seamlessly control a third robotic arm in conjunction with the natural arms, yielding smoother procedures, faster reactions, increased skill, reduced errors, and reduced manpower (fewer assistants). In fact, this is feasible because humans can naturally control their hands while carrying out other tasks using other parts of the body, e.g., one can manipulate objects with hands while speaking and walking.

Various hand-free interfaces have been developed to control the camera of laparoscopic surgery using head \cite{Kommu2007}, voice \cite{YulunWangGoletaDarrinUecker2002}, or switches activated by  fingers or feet \cite{Buess2000, Polet2008, Endex1993}. Most of these interfaces are intrinsically limited. For instance, voice commands may be affected by noise in operation room, and several persons cannot send commands together; turning the head away from the area of interest may affect hand movements; fingers' movements are coupled with the hand movements. Thus, the foot is the primary modality for inputs when the user's hands are busy \cite{Velloso2015, Alexander2012}. 

Foot controlled movements are easy to learn \cite{Simeone2014}, can be used in conjunction with hand movements \cite{Abdi2015,Abdi2016} and may provide similar or better skill as hand/voice controlled movements \cite{Podbielski2012,Footbetter2018}. Foot interfaces in commercial surgical robotic systems have been mainly used to control a laparoscopic camera \cite{Sackier1994, Voros2010, RoboLens2011}, and generally are footswitches or buttons placed closely on a planar base to move the camera at constant speeds in {zoom in/out, upward/downward, right/left} directions. Kawai et al. \cite{Kawai2014} used a pressure sensor sheet to record different foot patterns movements and locally control five degrees-of-freedoms (DOF) of a forceps, and Abdi et al. \cite{Doctoral2017} built an elastic-isometric four-DOF foot interface to control a robotic endoscope holder. These interfaces enable the movement to only a few discrete directions, i.e., it is difficult or not possible to command two or more DOFs simultaneously; and some require frequent visual checks to ensure that the foot is placed correctly, especially for novice users. These interfaces do not provide haptic feedback, which is required for fine control. Furthermore, the individual differences of foot operation are not considered in those interfaces.

In this paper we present a foot interface overcoming these limitations: a four-DOF parallel-serial hybrid mechanism with springs and force sensors. It allows intuitive control of a slave robotic arm in continuous directions and speeds, naturally matching the foot movements with dynamic force \& position feedbacks. The passive haptic feedback and automatic homing features of the interface relieve the user from visual checking of the foot positions. Moreover, the interface is adaptable to the specific movement patterns of different users so as to enable accurate control. 

The paper is organized as follows: Section II reveals the operation modes of the foot interface and selected four-DOF foot motions. Section III presents the proposed foot interface mechanism design, followed by the kinematics and statics modeling of the interface in Section IV. Section V introduces a user study to test the developed foot interface. The results exhibit the need to identify user-specific commands corresponding to their feet motion patterns. Section VI proposes and compares different mapping approaches for the user-specific motion patterns. Section VII summarizes the paper's contributions and discusses the interface's limitations.

\section{Feedback and sensing modules}

A passive compliant system with serial elastic feedback-sensing modules was selected to capture the continuous four-DOF natural movements of the foot with dynamic force feedback. Rich displacement and force feedback enables the operator to control robotic arms with enhanced intuitiveness, dexterity, and efficiency \cite{Mace2017a}. Elastic elements in robotic systems \cite{Passivelowerlimb2018,parallelelasticactuator2018} can help define the energy distribution of the system in the working range for desired functionality. 

\begin{figure*}[!t]
\centering
\includegraphics[width=0.8\textwidth]{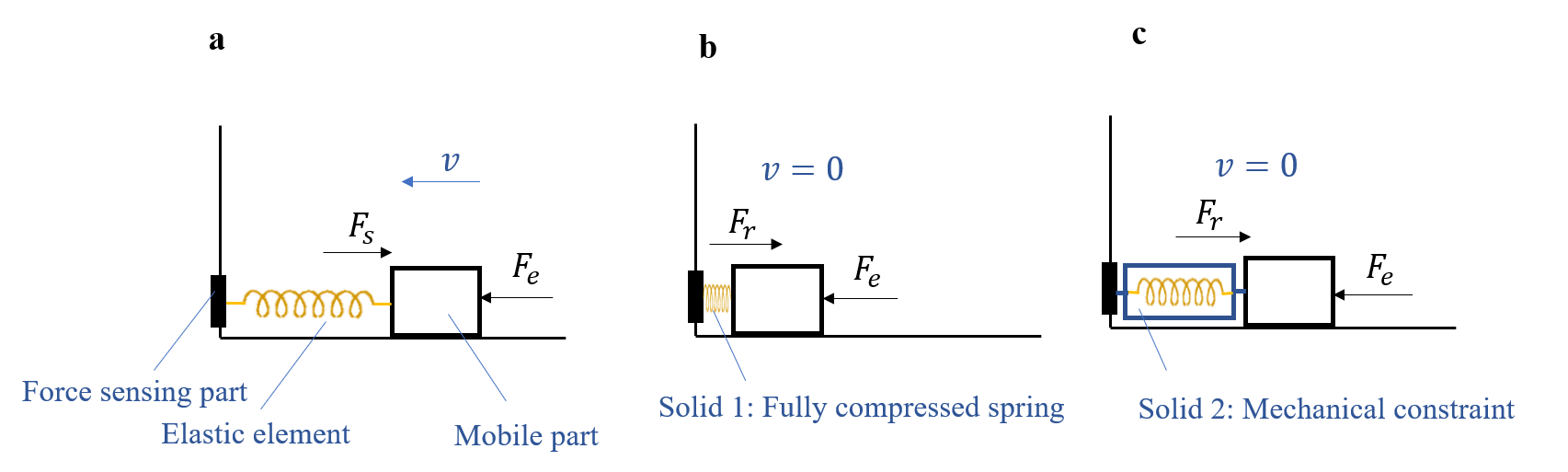}
\caption{Feedback and sensing module in (a) elastic and (b,c) isometric modes. $v$: velocity, $F_e$: force exerted by the foot, $F_s$: elastic force, $F_r$: reaction force.} 
\label{f:modeling}
\end{figure*} 

\subsection{Working Principle}
The feedback-sensing module consists of an elastic element and a force sensor \cite{gripAblePatent2016} as shown in Fig. \ref{f:modeling}. The force sensor, elastic element and the mobile part are connected serially with each other. The mobile part is activated by foot movements whilst elastic element is deformed under the force of the foot which is detected by the force sensor. The measured forces can then be used to calculate the deformations of the elastic module which are further used to calculate the position and orientation of the mobile part through kinematics. The whole foot operation can be separated into elastic mode (Fig. \ref{f:modeling}a) and isometric mode (Fig. \ref{f:modeling}b,c). Within the motion range of the elastic element, the foot is free to move (with reaction forces from the springs), and the position/force of the foot can be used as output of the interface to control a robotic arm; this control mode is elastic control mode. Once the elastic element reaches the elastic limit (e.g. the fully compressed compression spring in Fig. \ref{f:modeling}b) or mechanical constraint (Fig. \ref{f:modeling}c), the foot cannot further move beyond the corresponding boundaries, but the corresponding forces may still be changed by the user. This force signal provides isometric control mode. The external force $F_e$  from the foot can be derived from the spring force $F_s$ or reaction force $F_r$ via readings of the force sensor if the friction force is ignored. The transition between elastic and isometric modes enables sufficient proprioceptive information to the user and unlimited input range (depending only on the operator’s capability) which can be used to control the position/rate of the slave robot .

\subsection{Degrees of Freedom}
 
\begin{figure}[!t]
\centering
\includegraphics[width=\columnwidth]{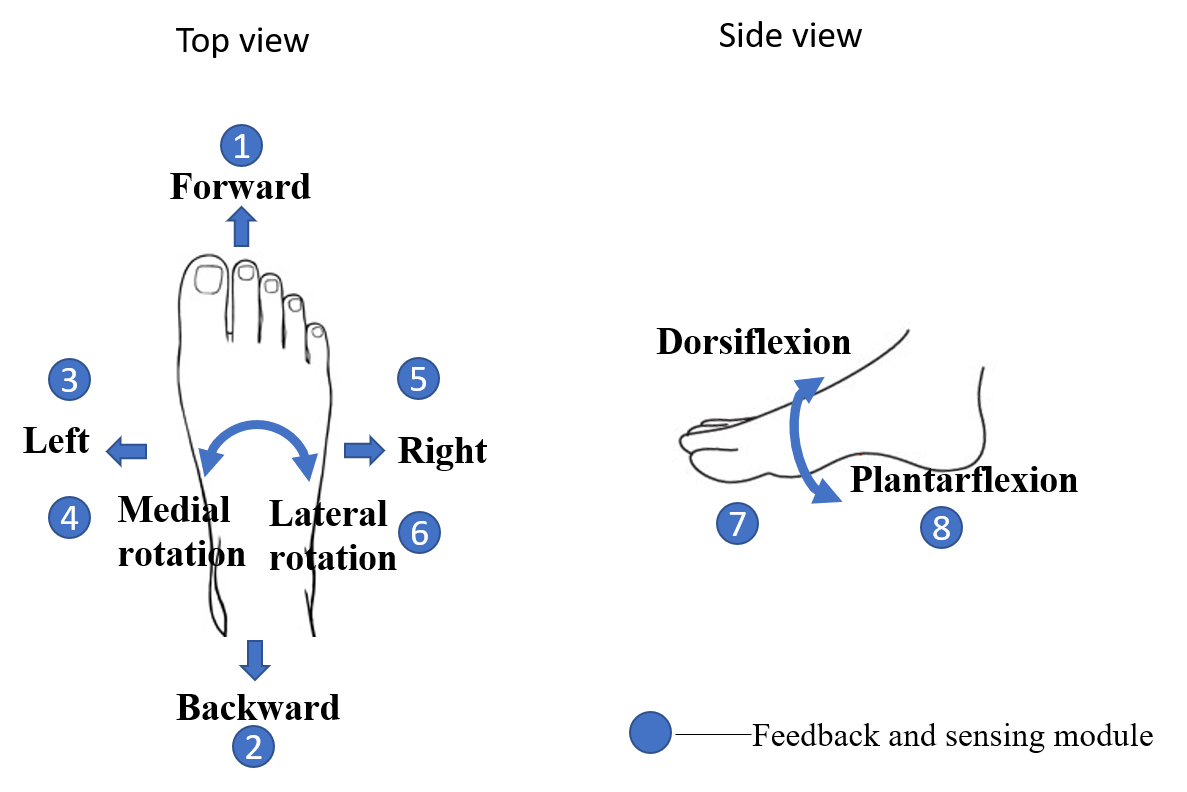}
\caption{Four-DOF foot motions enabled by the interface, and arrangement of feedback and sensing modules to measure these movements.}
\label{f:foot motion}
\end{figure}

Four-DOF specific foot motions were selected as the input to the foot interface system (Fig. \ref{f:foot motion}a): {\it i}) foot forward/backward movements (due to knee joint flexion/extension), {\it ii}) foot lateral movements (due to hip's abduction/adduction), {\it iii})foot lateral/medial axial rotation and {\it iv}) dorsiflexion/plantar flexion of the ankle. These natural foot movements can be carried out comfortably by most users. A relatively small workspace has been set to avoid uncomfortable operation and human fatigue (i.e. The above mentioned motions {\it i}) and {\it ii}) are limited to 2cm; motions {\it iii}) and {\it iv}) are within $12.5^o$ and $10^o$). 

Eight feedback and sensing modules are located around the foot as shown in Fig. \ref{f:foot motion} to collect the input signals from the foot motions. The first six feedback and sensing modules can detect three-DOF foot motions in the horizontal plane. They are parallel connected to the foot to provide dynamic and continuous elastic feedback. Two additional feedback and sensing modules are located under the sole and heel to collect force signal of the fourth DOF foot motion in dorsiflexion/plantar flexion. The details of the design and modeling of the foot interface mechanism are described in the next two sections.

\section{Interface design}
\begin{figure*}[!t]
\centering
\includegraphics[width=\textwidth]{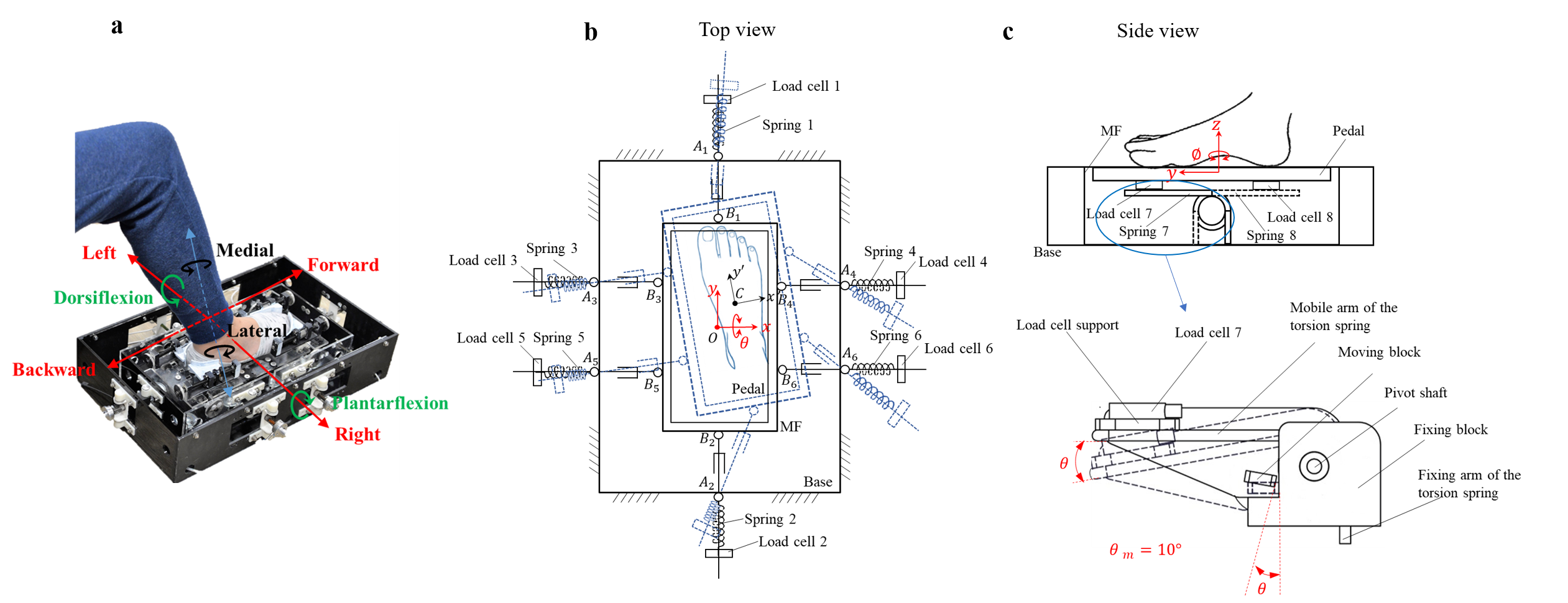}
\caption{Passive four-DOF foot interface. (a) The interface prototype controlled by a user's foot. (b) The schematic diagram in top view at initial home position (blue dotted lines show a random position of the MF) and (c) side view with mechanical motion limits.}
\label{f:system}
\end{figure*} 
Fig. \ref{f:system} shows the foot interface prototype and the schematic diagrams. The foot is fixed to a foot pedal which is serially connected with a mobile frame (MF) through a pivot shaft, torsion springs, and load cells. The MF connects to the base through a series of linear guides, compression springs, and load cells, forming a parallel structure. The whole system is essentially a parallel-serial hybrid mechanism controlled by foot. The parallel mechanism provides a simple and compact structure with low inertia through closed-loop kinematics; meanwhile, the serially connect pedal decouples the dorsiflexion/plantar flexion movements of the foot from other horizontal movements within a relatively large workspace.

The elastic mode of the system is achieved through the elastic network with eight springs providing dynamic real-time passive force \& position feedback. In addition, these springs are carefully arranged for a singularity-free workspace with a neutral central home position (global minimum elastic energy), as sketched with solid lines in Fig. \ref{f:system}b,c. When the operator finishes operation movements and releases the pedal, the interface returns to the home position automatically (assuming zero friction), providing a resting posture for the foot and enabling a quick start for the next operations, without the need of a visual check. The usage of force sensors instead of position sensors collects force information of the operator’s foot within and even beyond the geometric workspace, enabling 9 transition between elastic and isometric modes. 

\begin{figure*}[!t]
\centering
\includegraphics[width=0.95\textwidth]{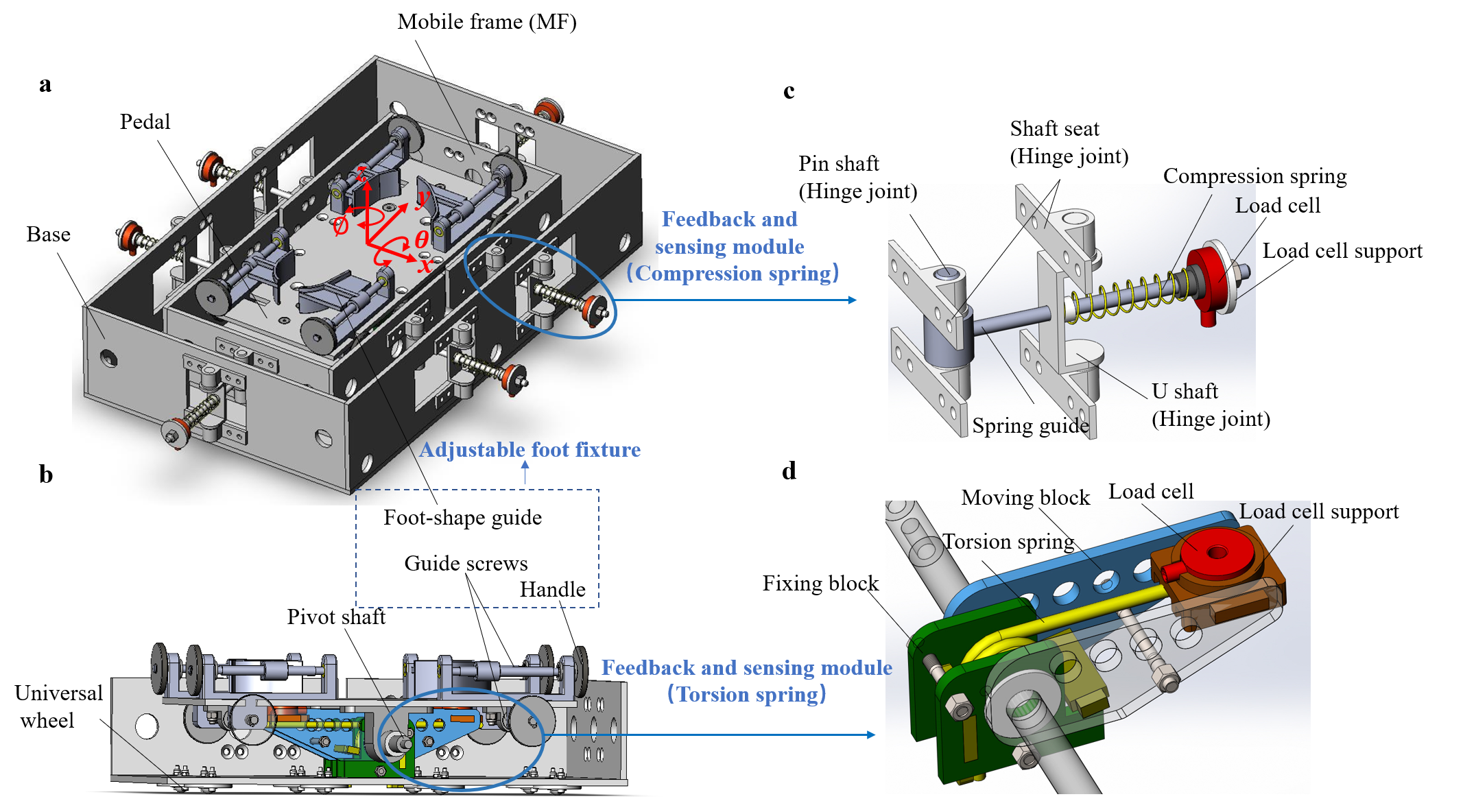}
\caption{3D system mechanical structure of the foot interface in (a) perspective top, (b) open side views and zoom views of (c) compression spring feedback and sensing module, (d) torsion spring feedback and sensing module.}
\label{f:3D}
\end{figure*} 

The 3D model of the foot interface is detailed in Fig. \ref{f:3D}. The interface includes a base which is fixed on the ground, a mobile frame (MF), a pedal with adjustable foot fixture and the feedback and sensing modules with springs and force sensors. The base and the MF have a parallel kinematics structure and are connected by six feedback and sensing modules of the compression spring (Fig. \ref{f:3D}c) with hinge joints on both sides. The MF is the input component which can slide in the horizontal plane in two translations $x,y$ and one rotation $\phi$ (2T1R) within the base. To reduce the friction and inertia, eight universal wheels are mounted at the bottom of the MF (Fig. \ref{f:3D}b), which transform sliding friction to rolling friction in order to minimize a user's fatigue. The pedal plate for the foot is serially mounted in the MF as an extension with a pivot shaft and two feedback and sensing modules of the torsion spring (Fig. \ref{f:3D}d), acting as a second input component for pitch rotation $\theta$. 

The pitch rotation $\theta$ of the pedal and movements  $x,y,\phi$ of MF are decoupled so they do not affect each other. The potential motion coupling problem between forward/backward motion and dorsiflexion/plantar flexion of the ankle was minimized by placing the pitch pivot shaft at a lower position between the height of the spring guide and motion surface. The driving force from the human foot acting on the pivot shaft, the reaction forces of the springs, and the friction force (low) are generally counterbalanced. 

An adjustable foot fixture mechanism is mounted on the foot pedal plate enabling comfortable but rigid fixation. This fixture is composed of four 3D-printed, foot-shaped guides that can fit well different human feet. Each guide block connects with two guide screws, enabling four directions adjustments independently or in tandem, for feet of size 35 to 46 in Europe standard. The length and width can be easily adjusted by twisting the handles.

\begin{table}[!t]
\begin{center}
\caption{Springs' specification}
\label{tab:spring}
\begin{tabular}{|l|l||l|l|}
\hline
compression spring & & torsion spring & \\\hline
stiffness& $0.02\,N/cm$ & stiffness & $46.3 \,N\,cm/^ {\circ}$ \\\hline 
free length& $6\,cm$& position angle &$90^{\circ}$\\\hline
initial length &$3.2\, cm$& pre-tension& $0 N$\\\hline
fully compressed length &$1.2\,cm$&operating angle&$26^ {\circ}$\\\hline
\end{tabular}
\end{center}
\end{table}

The six compression springs and a pair of torsion springs form a spring network with four DOFs. Springs of stiffness 0.01, 0.02, 0.03, 0.05, 0.1 N/cm were considered, and the 0.02 N/cm compression spring was selected as not too hard to press while providing sufficient haptic feedback. Compression springs were preferred than extension springs, as they enable force measurement even when fully compressed and are simple to assemble. The selected spring specifications are listed in Table \ref{tab:spring}. A pair of torsion springs with $90^o$ position angle and $46.3 N \, cm/^o$ spring stiffness was selected for the pitch rotation DOF mounted symmetrically but in reverse directions on the pivot rotation shaft which support the pedal plate. For each torsion spring, the fixed arm is integrated to the MF via a fixing block, whereas the mobile arm and moving block rotate with the pedal plate. The twisting angle was mechanically limited to $\pm10^o$ which is reached once the moving block touching the vertical plane of fixing block (refer to Fig. \ref{f:system}c enlarged view). The spring mechanism provides real-time force feedback changing monotonically with the foot displacement. Each spring is located outside the base mounting in serial with the corresponding force sensor (LW1025-25 from Interface, Inc., USA). As will be analyzed in Fig. \ref{f:energy} of Section \ref{s:modeling}, this avoids singularities inside the workspace and defines an automatic home position.

The applied forces by the operator are translated into electrical signals through eight force sensors. To ensure that the springs can effectively transmit forces to load cells in any poses of the pedal, a 5.6N pre-compression force is applied for each compression spring at home position. While the pedal’s movements are limited by the deflection range of compression spring (for DOFs in 2T1R) and mechanical constraint (for DOF in pitch rotation), the force detection range is not restrained, i.e., once a compression spring is fully compressed, or the pitch rotation reaches the limit angle, the isometric force just builds up and is still measured by the load cell. 

\section{Modeling}	\label{s:modeling} 
\subsection{Kinematics}	\label{ss:kinematics}
The kinematics of the three DOFs of 2T1R in the horizontal plane (without tilting the paddle) can be regarded as a 6-RPR planar parallel mechanism as shown in Fig. \ref{f:system}b. The MF is constrained and connected to the base via six spring guides with hinge joints on both sides (represented as points $A_i$ on the base and $B_i$ on the MF). A fixed base reference frame \{O-$xy$\} and mobile reference frame \{C-$x'y'$\} are assigned to the centroid of the base and the MF square plane, respectively. 
 
\begin{figure*}[!t]
\centering
\includegraphics[width=0.9\textwidth]{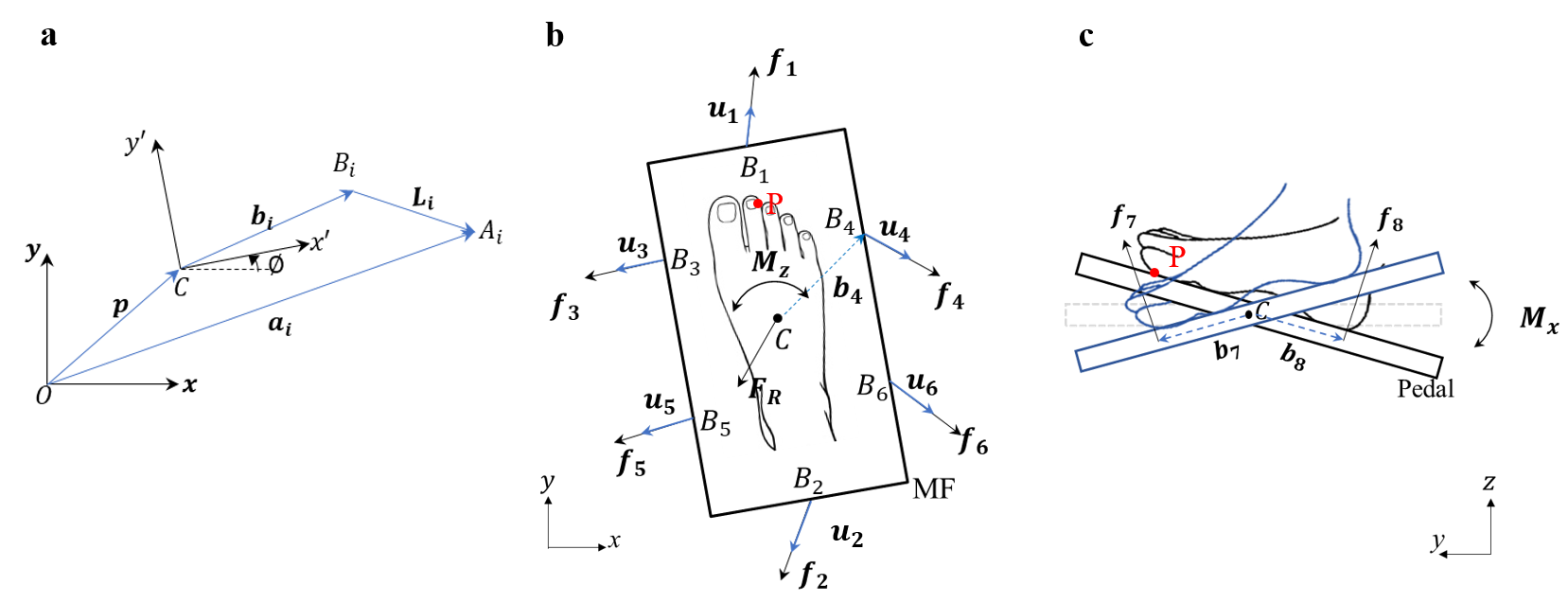}
\caption{(a) Kinematic model of \textit{i}th compression spring and (b) statics forces in $x$-$y$ horizontal plane and (c) pitch rotation DOF.}
\label{f:kinematics}
\end{figure*}
The position vector of point $A_i$ is defined by the vector $\mathbf{a}_i \equiv [a_{ix}, a_{iy}]^T$ expressed in the fixed frame \{O\} whereas the position vector of point $B_i$ is defined as $\mathbf{b}_i \equiv [b_{ix},b_{iy}]^T$ in the mobile frame \{C\}. They can be represented using $a,b,a',b'$, the lengths and widths of base the and the MF square, and $c,c'$, the distance between the laterally located linear spring guides, equal to lengths $A_3A_5$ and $B_3B_5$. The position of the MF reference frame \{C\} relative to the base reference frame \{O\} is defined by the vector $\mathbf{p} = [x,y]^T$ connecting O to C, while its orientation is defined by the angle $\phi$ between the $\widehat{x_O}$ axis of the \{O\}  and the $\widehat{x'_{C}}$ axis of the \{C\}. The length of \(i\)th spring guide between the attach points of base $A_i$ and the MF $B_i$ is denoted as ${L}_i$. Finally, let $\mathbf{u}_i \equiv \overrightarrow{B_iA_i}/{L_i}$ denote the unit vector along \(i\)th spring guide, and \(\mathbf{L}_i \equiv L_i\mathbf{u}_i\).

Using these definitions, the closed mechanical chain of Fig. \ref{f:kinematics}a is:
\begin{equation}
\label{e:closedChain}
\mathbf{L}_i = \mathbf {a}_i  - \mathbf {p} -\mathbf{R} \, \mathbf{b}_i \,, \quad 
\mathbf{R} \equiv 
\left[ \begin{array}{cc}
\cos\phi & -\sin\phi\\
\sin\phi & \cos\phi
\end{array} \right]
\!,
\end{equation}
where $\mathbf{R}(\phi)$ is the rotation matrix from the fixed frame \{O\} to the mobile frame \{C\}. For this parallel mechanism, the  {\it inverse kinematics}, i.e., calculating the guide lengths ${L}_i$ as a function of the pose translation $\mathbf{p} = [x,y]^{T}$ and rotation $\phi$ can be computed from the closure constraints:
\begin{equation}
\label{e:closure}
\begin{aligned}
L_i^2&=(x + b_{ix} \cos\phi - b_{iy} \sin\phi - a_{ix})^2 \\ &+(y + b_{ix} \sin\phi + b_{iy} \cos\phi - a_{iy})^2\,.
\end{aligned}
\end{equation}
For the {\it forward kinematics}, given $L_i, i = 1,2...6$, $\mathbf{X} = [x,y,\phi]^T$ can be derived by solving the six scalar equations (\ref{e:closure}), yielding
\begin{eqnarray}
\begin{aligned}
\label{e:forwardkinematics}
\phi &= \arcsin \frac{E}{2P} \\
x &= \frac{F(a' \sqrt{4P^2-E^2}-2aP)-2bEG}{4Q \sqrt{4P^2-E^2}-8PM}  \\
y &= \frac{2G(b' \sqrt{4P^2-E^2}-2bP)+a'EF}{4Q \sqrt{4P^2-E^2}-8PM} \quad,
\end{aligned}
\end{eqnarray}
where $E \equiv L_4^2 + L_5^2 - L_3^2 - L_6^2$, $F \equiv L_3^2 + L_5^2 - L_4^2 - L_6^2$, $G \equiv L_2^2 - L_1^2$, $M \equiv ab + a'b'$, $Q \equiv ab' + a' b$ and  $P \equiv bc' - b'c$. The magnitude of guide length \(L_i\) is derived from delta compression spring forces \(f_i\) via Hooke's Law:
\begin{equation}
\label{e:hooke's law }
L_i = L_{0i} + \frac{f_i}{k_i},\, f_i = f_{ci} - f_{0i},\,i = 1,2,...6,
\end{equation}
where $k_i$ is the compression spring stiffness constant of \(i\)th spring. \(L_{0i}\), \(f_{0i}\) is the original length and pretension force  of \(i\)th spring guide and spring at home position. \(f_{ci} \) is \(i\)th load cell's reading. The pitch rotation angle is
\begin{equation}
\label{e:forwardkinematics_4dof}
\theta=
\frac{M_x} {K_p}\,,
\end{equation}
where \(M_x\), \(K_p\) are the rotation torque and stiffness constant in pitch DOF which will be derived in eq.(\ref{e:statics_4dof}). 
 
\subsection{Statics}
\paragraph{Elastic statics}
Gravity can be neglected in the horizontal plane of Fig. \ref{f:kinematics}b. The resultant force and torque of $\mathbf F_R$ and $\mathbf M_z$  can be computed from delta compression spring forces  \(\mathbf {f}_i = f_i \mathbf {u}_i,\,i =1,2,...6\).  The statics equations are:
\begin{equation}
\label{e:statics}
\begin{aligned}
&\mathbf F_R =\sum_1^6\mathbf f_i\,  \quad \\
&\mathbf M_z = \sum_1^6\mathbf f_i \times \mathbf{R} \, \mathbf{b}_i = \sum_1^6 \mathrm{det} (\mathbf{f}_i,\mathbf{R} \, \mathbf{b}_i) \,,
\end{aligned}
\end{equation} 
which can be written in matrix form as
\begin{equation}
\begin{aligned}
\label{e:structurematrix}
\mathbf{W}_R &= \mathbf{J} \, \mathbf{f}\,, \quad  \,\mathbf f \equiv \left[\!\! \begin{array}{c} \mathbf{f}_1 \\ \vdots \\ \mathbf{f}_6 \end{array} \!\! \right], \quad \mathbf{W}_R \equiv \left[\!\! \begin{array}{c} \mathbf{F}_{xR} \\ \mathbf{F}_{yR} \\ \mathbf{M}_z \end{array} \!\! \right],  \quad \\
\mathbf{J} \! &= \! \left[\!\! \begin{array}{c} \mathbf{u}_1\,\,, \cdots \,\,\,, \mathbf{u}_6 \,\, \\ \mathrm{det}(\mathbf{u}_1 \!\,, \,\mathbf{R}   \, \mathbf{b}_1) \,\,,\cdots \,\,, \mathrm{det}(\mathbf{u}_6 \!\, ,\,\mathbf{R} \, \mathbf{b}_6) \end{array} \!\! \right]\!,  
\end{aligned}
\end{equation}
where \(\mathbf{J}\) is the structure matrix of the planar parallel structure. The stiffness matrix \(\mathbf{K}\) in motion workspace can be found by taking derivatives of  \(\mathbf{W}_R\) with respect to \(\mathbf{X}\):
\begin{eqnarray}
\label{e:stiffness}
\mathbf{K} \!\!&=&\!\! \frac{d\mathbf{W}_R}{d\mathbf{X}} =\,
\mathbf{J} \, \frac{d\mathbf{f}}{d\mathbf{X}} + \frac{d\mathbf{J}}{d\mathbf{X}} \, \mathbf{f} \, = \, \mathbf{J}\mathbf{C} \mathbf{J}^T \! +  \frac{d\mathbf{J}}{d\mathbf{X}} \, \mathbf{f}\,, \quad \nonumber \\ 
\mathbf{C} \!\! & \!\! \equiv& diag(k_1 ... k_6) \, .
\end{eqnarray}
The torque for pitch rotation \(M_x\) around center point \(C\) is obtained through the following equation: 
\begin{equation}
\label{e:statics_4dof}
M_x =  f_7 \, b_7\, -\,f_8 \, b_8 \, , \quad
K_p=
\begin{cases}
 k_7 & \text{if $M_x>0$}  \\
 k_8 & \text{if $M_x<0$} 
\end{cases}\, 
,
\end{equation}
where \(f_7\) and \(f_8\) are recorded by load cells 7 and 8 placed under the sole and heel, respectively. Theoretically, there is no pre-tension force for the two torsion springs in the balanced state. The force magnitude of \(f_7\) and \(f_8\) directly reflect the input force change (Fig. \ref{f:kinematics}c). \(b_7\) and \(b_8\) are the arm lengths from pedal plate center \(C\) to positions of load cells 7 and 8, which can be modified according to the operators' habitual posture. The stiffness \(K_p\) in this dof is directly reflected in \(k_7, k_8\), the stiffness constants of torsion springs 7 and 8.

The \(\mathbf J_s\) and \(\mathbf{K}_s\) of the four-DOF structure can be rewritten as below:
\begin{eqnarray}
\label{e:totalmatrix}
\mathbf {W} \!\!\!\!&=&\!\!\! \mathbf{J}_s\mathbf{f}\, , \quad
\mathbf{f} = \!\left[\!\! \begin{array}{c} \mathbf {f}_1 \\ \vdots \\ \mathbf{f}_8 \end{array} \!\! \right]\!, \quad 
\mathbf{W} = \!\left[\!\! \begin{array}{c}\mathbf{F}_{xR} \\ \mathbf{F}_{yR} \\ \mathbf{M}_z \\ \mathbf{M}_x\end{array} \!\! \right]\!,  \quad  \\
\mathbf{J}_s \!\!\!\!\! &=& \!\!\!\!\! \left[\!\! \begin{array}{cc} \mathbf{J} \!\!& \!\!\mathbf{0} \\ \mathbf{0} \!\!& \!\! \mathbf{J}_p \end{array} \!\! \right]\!, \quad  
\mathbf{J}_p = \left[b_7 \,, \,-b_8\right]\,, \quad 
\mathbf{K}_s \! = \! \left[\!\! \begin{array}{cc} \mathbf{K} \!\! & \!\! \mathbf{0} \\ \mathbf{0} \!\! & \!\! \mathbf{K}_p \end{array} \!\! \right]. \nonumber
\end{eqnarray}
where the resultant wrench $\mathbf{W}$ is feedback to the human. From static equilibrium, the sum of external force and moments $\mathbf{W}_e$ exerted on the pedal equals the resultant external wrench $\mathbf{W}$ exerted on the human foot.

\paragraph{Isometric statics}
When the pedal reaches a workspace’s boundary, the system is in isometric mode. The force/torque can continue to increase as is recorded by load cells coupled with the fully compressed compression spring(s) or the mechanically constrained torsion spring.

\begin{figure*}[!t]
\centering
\includegraphics[width=0.95\textwidth]{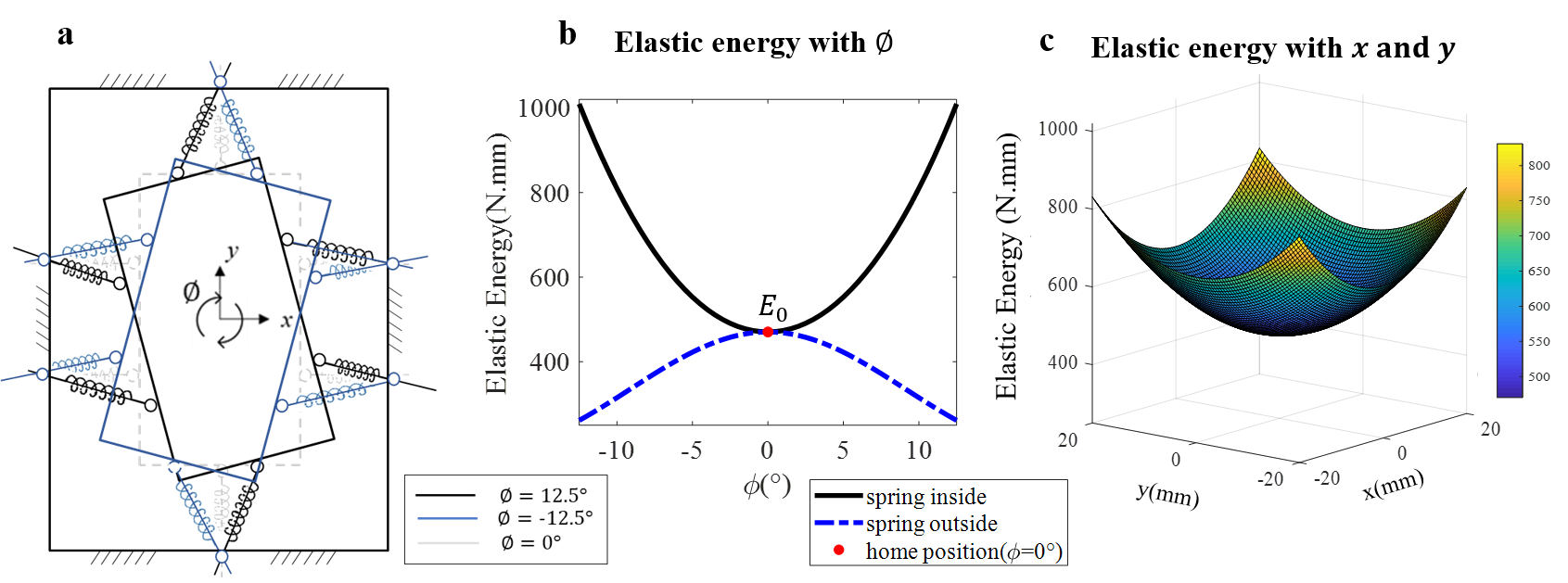}
\caption{(a) Initial design with springs inside the base, between the base and MF. (b,c) show how elastic energy depends on yaw the rotation angle $\phi$ (b) and on $\{x,y\}$ (c).}
\label{f:energy}
\end{figure*}

\paragraph{Singularities}
Fig. \ref{f:energy}a illustrates a previous design with compression springs between the base and MF. This design yields two singular configurations (Fig. \ref{f:energy}a, black and blue lines) at the extreme yaw rotations, where the pedal remains twisted, making the pedal's movement uncontrollable. This issue was addressed by placing the compression springs outside of the base (Fig. \ref{f:system}b). This changes a compressing force exerted on the MF to a pulling force, thus making the springs free to take any orientation independently on the other connections. This structure avoids the mechanical constraints that result from using bars connecting the base and MF and the resulting singularities at local energy minima (Fig. \ref{f:energy}b, blue line). The elastic energy has a unique global minimum value at home position (Fig. \ref{f:energy}b, black line and Fig. \ref{f:energy}c). 

\section{Analysis of aiming error}\label{s:error}
\subsection{Experiment}	\label{ss:experiment}
An experiment was conducted with ten subjects (27.3\(\pm\)2.2 years old, right foot dominant, 4 female) to study the precision with which they can control the interface in given directions using their right foot. The experiment was approved by the Institutional Review Board (IRB) of Nanyang Technological University (IRB-2018-05-051).

Each subject was seated comfortably on a chair in front of a table, and the foot interface was placed under the table at the home position with the pedal horizontally located at the center of the base. The subjects were asked to step their right feet on the pedal and adjust the positions of the fixture blocks to match their specific feet sizes.

The {\it task} was to move the pedal from the home position along multiple specified directions to the boundary of the workspace. Once at the boundary, the pedal should be held for one second  and returned back to the home position. The task was demonstrated to the subjects by the experimenter prior to data collection. The subjects were also informed of the desired target directions. No visual feedback of the foot posture was provided during the experiment. In this way, we could observe feedforward foot movements corresponding to any desired motion directions. Movements were carried out in single and diagonal directions:

\paragraph{Single directions} 
Each subject started at home position. Three trials were conducted in each of the following eight directions: forward (F), backward (B), left (L), right (R), toe up rotation (TU), toe down rotation (TD), left torsion (LT), and right torsion (RT) in this order. This procedure was continuous without pause until all the \(3\times8 = 24\) centre-out and back movements were completed. Then, the subject lifted down his/her foot and took a 30 seconds' break. Another group of 24 trials was repeated after that.
\paragraph{Common diagonal directions} 
Each subject started at the home position. Three trials were conducted in the twelve common diagonal directions, which are the combination of the two single Cartesian directions, in the order of \{left \& forward (LF), right \& forward (RF), left \& backward (LB), right \& backward (RB), left \& toe up (LTU), right \& toe up (RTU), left \& toe down (LTD), right \& toe down (RTD), forward \& toe up (FTU), backward \& toe up (BTU), forward \& toe down (FTD), backward \& toe down (BTD)\}. \(3\times12 = 36\)  trials were conducted in diagonal directions.

\subsection{Data Analysis}  
Foot force data from load cells were recorded at 50 Hz and smoothed offline by using a moving average filter with window size of 9. They were mapped to the pose vector $\boldsymbol{D} = [x,y,\phi,\theta]^T$ of the center point C on the pedal using the forward kinematics Eqs.  (\ref{e:forwardkinematics}, \ref{e:forwardkinematics_4dof}). 

The actual initial position of the foot for each consecutive operation was considered as the calibrated home position. To compare different components of the pose, a point P on the pedal, $\mathbf P = [0, d, 0, 0]$ (with $d=11.5cm$)  with respect to \{C\}, was selected as a reference point (see Fig. \ref{f:kinematics}). The four-DOF position change of point P with respect to home position is represented as $\mathbf {P}_{delta} = [P_x,P_y,P_\phi, P_\theta]$ (section \ref{ss:experiment}), where $P_x = x$, $P_y = y$, $P^2_\phi = (d\sin{\phi})^2 + (d-d\cos\phi)^2$, $P_\theta = d\sin\theta$, with the range $P_x, P_y, P_\theta \in [-2, 2](cm)$, $P_\phi \in [-2.5,2.5](cm)$. A scaling factor $s = 2/2.5$ was used for $P_\phi$ to bring the two angles in the same range. The data of vector \(\mathbf P\) were then filtered to remove the static position data using a resultant velocity threshold of 0.005m/s.

{\it Foot-path error} was used to quantify the performance in the trials:
\begin{equation}
E = \frac{1}{N}\sum_{i=1}^{N} ||\mathbf{e}_i || \,, \,\,\,\, \mathbf{e}_i \equiv \mathbf{p}_i - \hat{\mathbf{p}}_i = [e_{ix}, e_{iy}, e_{i\phi}, e_{i\theta}]\,,
\end{equation}
where \(\mathbf{p}_i\) is the real trajectory points, $\hat{\mathbf{p}}_i$ the projection of \(\mathbf{p}_i\)  on the desired path, and $N$ is the total number of samples for foot movements in each direction. 

\subsection{Results}
\begin{figure*}[ht]
\begin{center} 
\includegraphics[width=0.95\textwidth]{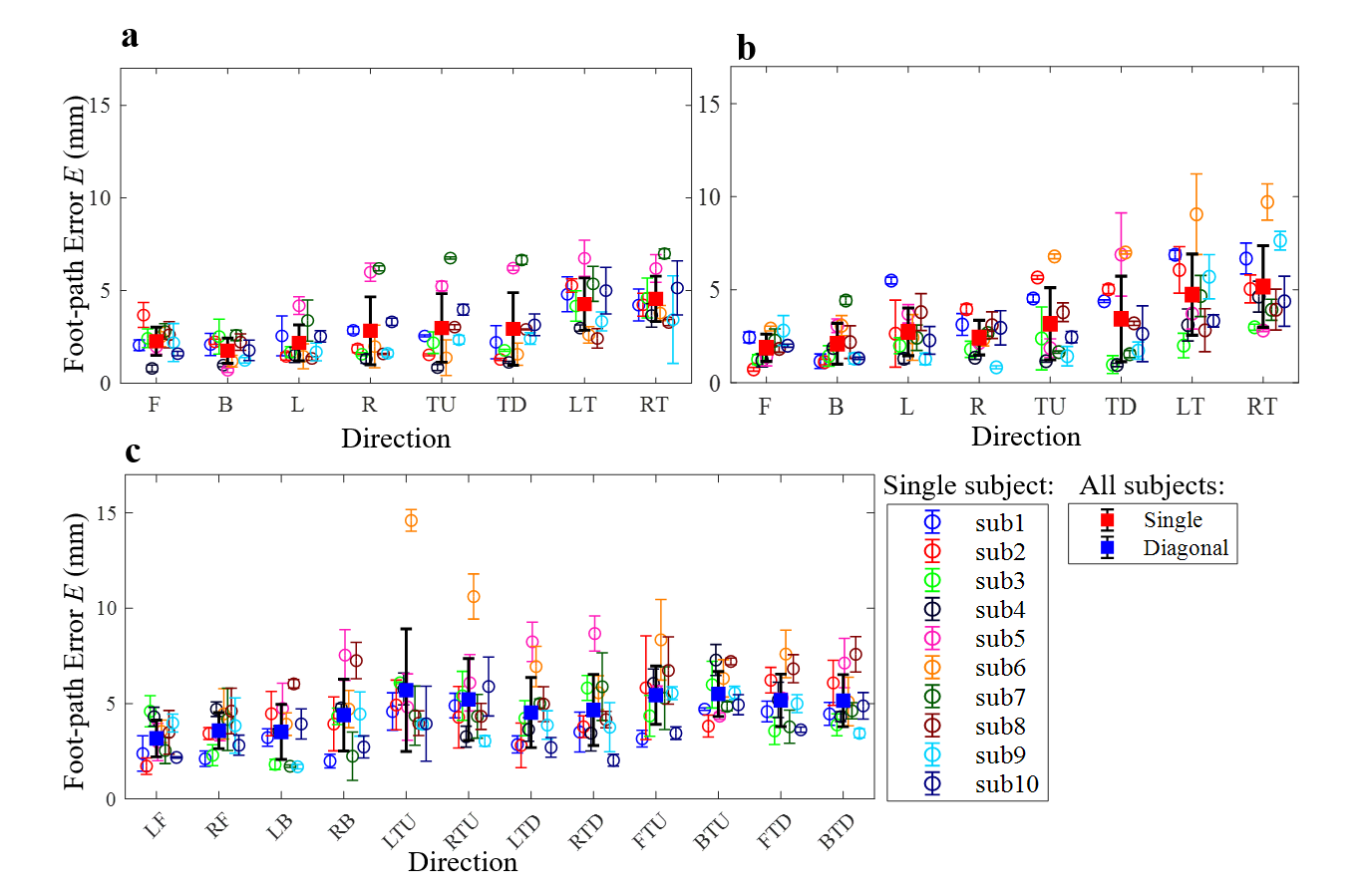}
\caption{Foot-path error and its standard deviation over subjects and consecutive trials for each subject on (a) data set 1 of single-direction trials, (b) data set 2 of single-direction trial and (c) data set 3 of diagonal-direction trials.}
\label{fig:Etotal_trial}
\end{center}
\end{figure*}

Fig. \ref{fig:Etotal_trial} shows the foot-path error \(E\) over the ten subjects in various directions and the variability over the consecutive trials of each subject. We see that in the single directions (Figs.\ref{fig:Etotal_trial}a,b) LT and RT have a larger error relatively to the other directions, and F/B have the least error. Diagonal directions (Fig. \ref{fig:Etotal_trial}c) that combine translations in $x-y$ planes have less error then the diagonal motions involving rotation and translation in $x-\theta$, $y-\theta$ planes. In most cases, the deviation is large between subjects. And the deviation is relatively small in the trials of each subject. In addition, the smoothness in the two data sets of single direction trials and the data set of diagonal direction trials were analyzed using the spectral metrics \cite{Balasubramanian2015} which can be used to assess learning \cite{Balasubramanian2012}. There was no significant difference between the two single data sets (T-test, p$=$0.4), meaning that there was no learning effect between these two periods. However, there is a different level of smoothness (T-test,  p$<$0.01), indicating smoother movements in single axis directions as in diagonal ones.

\section{Subject-specific control patterns}
The forces measured by the interface will reflect the operator’s foot motion intention and be used to control a device with four-DOF. What is required is a mapping: 
\begin{equation}
\left[\begin{array}{c} f_1 \\ \vdots \\ f_8 \end{array} \right] \rightarrow  \left[\begin{array}{c} x \\ y \\ \phi \\ \theta \end{array} \right]
\end{equation}
from the eight load cells' force signals to a movement in the four DOFs of $\mathbf{D} \equiv \{x, y, \phi, \theta\}$, reflecting the pose of center point C of the pedal under base frame \{O\} (Figs. \ref{f:system}b,c). An obvious solution consists of using the forward kinematics Eqs.(\ref{e:forwardkinematics}, \ref{e:forwardkinematics_4dof}) which directly reflect the foot motion patterns. 

As discussed in the last section, the foot motions’ performances are different depending on the direction and subject. However, the variability over consecutive trials of a single subject is relatively low as can be seen in Fig. \ref{fig:Etotal_trial}. Therefore, an alternative idea for the mapping consists of identifying subjects- and directions- specific force patterns that can be used as commands to control a four-DOF device by foot. The independent component analysis (ICA) can separate mixing signals into simpler components. This method was used to find the foot motion pattern of the subject and obtained a superior performance than kinematics modeling. 

The force data \(\mathbf{f}_n\) is the input to the ICA model, which is the z-score normalization form of delta force changes of load cells’ readings  \(\mathbf{f}\) (from load cells real-time readings \(\mathbf{f}_c\) subtracting the spring pre-tensions \(\mathbf{f}_0\) at the home position). The FastICA algorithm \cite{HYVARINEN2000411} is then applied onto this data to derive a subject specific ICA model, as
\begin{eqnarray}
\label{e:ICA}
\mathbf{D}_n \!\!\! &=& \!\!\!\mathbf{T} \, \mathbf{f}_n \, , \quad 
\mathbf{D} _n \equiv \! \left[ \begin{array}{c} x_n \\ y_n \\ \phi_n \\ \theta_n \end{array} \right]\!, \nonumber \quad 
\mathbf{T} \equiv\! \left [\begin{array}{c} \mathbf{T}_x\\\mathbf{T}_y\\\mathbf{T}_\phi\\\mathbf{T}_\theta\end{array} \right]\!, \quad \\
\mathbf {f}_n \!\!\! &=& \!\!\! \frac{\mathbf {f}-\mathbf {\bar{f}}} {\mathbf {\sigma}}\,, \quad
\mathbf{f} \equiv \left[\begin{array}{c} f_{c1}-f_{01} \\  f_{c2} -f_{02}\\. \\.\\. \\ f_{c8}-f_{08} \end{array} \right]\!, 
\end{eqnarray}
where \(\mathbf{T}\) is a \(4\times 8\) mapping matrix from one subject's force data.  Its components \(\mathbf{T}_x,\mathbf{T}_y,\mathbf{T}_\phi,\mathbf{T}_\theta\) are eight-dimension arrays for each DOF, which are derived from single axis Cartesian motion data of L \& R, F \& B, LT \& RT, TU \& TD respectively. They reflect the new basis of four DOFs for the specific subject in the combination of eight load cell readings.

Once a subject-specific model is built using the calibration procedure described in the single direction task of section \ref{ss:experiment}, the subject-specific patterns are used to predict further motor commands matching one’s motion intention, i.e., mapping the force \(\mathbf{f}\) to motion command \(\mathbf{D}\). To verify this method, we analyse the three sets of data collected in section \ref{ss:experiment} that are used for modeling (single axis direction data set 1) and testing (single axis direction data set 2 and diagonal axis direction data). For comparison purposes, the results from using the kinematics and ICA methods are converted to the same range \([-1,1]\) through a min-max normalization. The original positions are calibrated to the zero for each consecutive set of data.
\begin{figure*}[!t]
\centering
\includegraphics[width=1\textwidth]{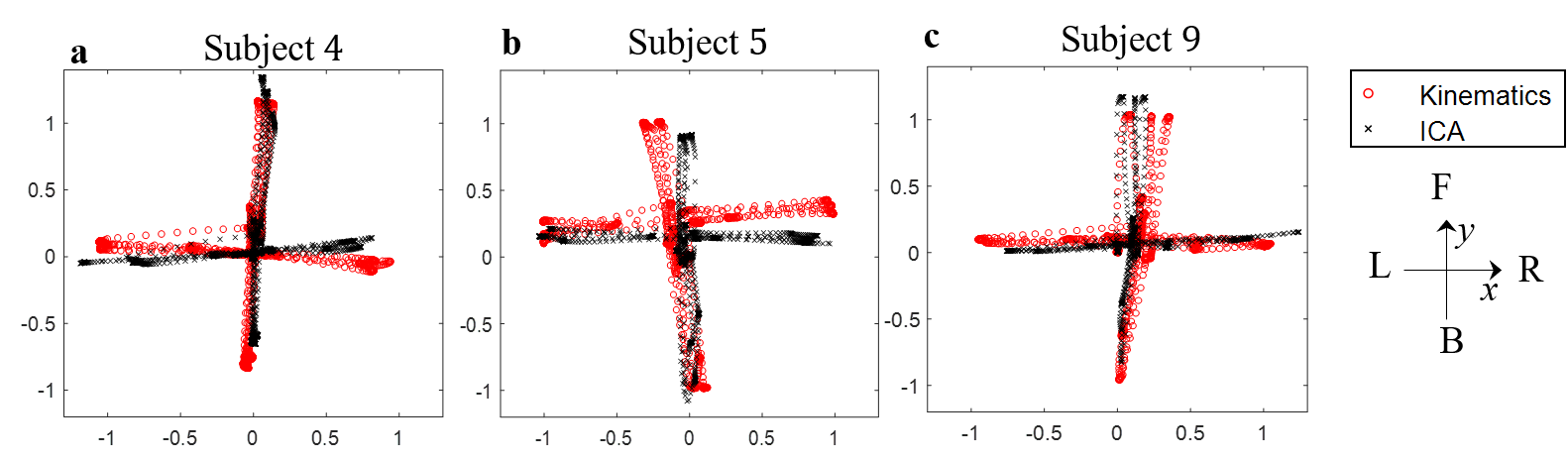}
\caption{Comparison of foot motion path in directions F, B, L, R of three representative subjects for single Cartesian modeling data using kinematic transformation and ICA.}
\label{f:3subjects}
\end{figure*}
Fig. \ref{f:3subjects} illustrates the effect of using for transformation subject- and direction-specific ICA compared to kinematics modeling, with three subjects' foot trajectories in the $x-y$ plane. It is plotted by the result of \(\mathbf{D}_{n,xy}\) = \([x_n, y_n]^{T}\) with modeling data in directions of F, B, L, R, with three trials for each direction. The kinematics model (red traces) corresponding to the actual human foot motions exhibits different patterns over subjects. In contrast, the results of ICA modeling yield a subject- and direction- specific mapping minimizing the foot-path error and performance variability of users.
\begin{figure*}[!t]
\centering
\includegraphics[width=0.95\textwidth]{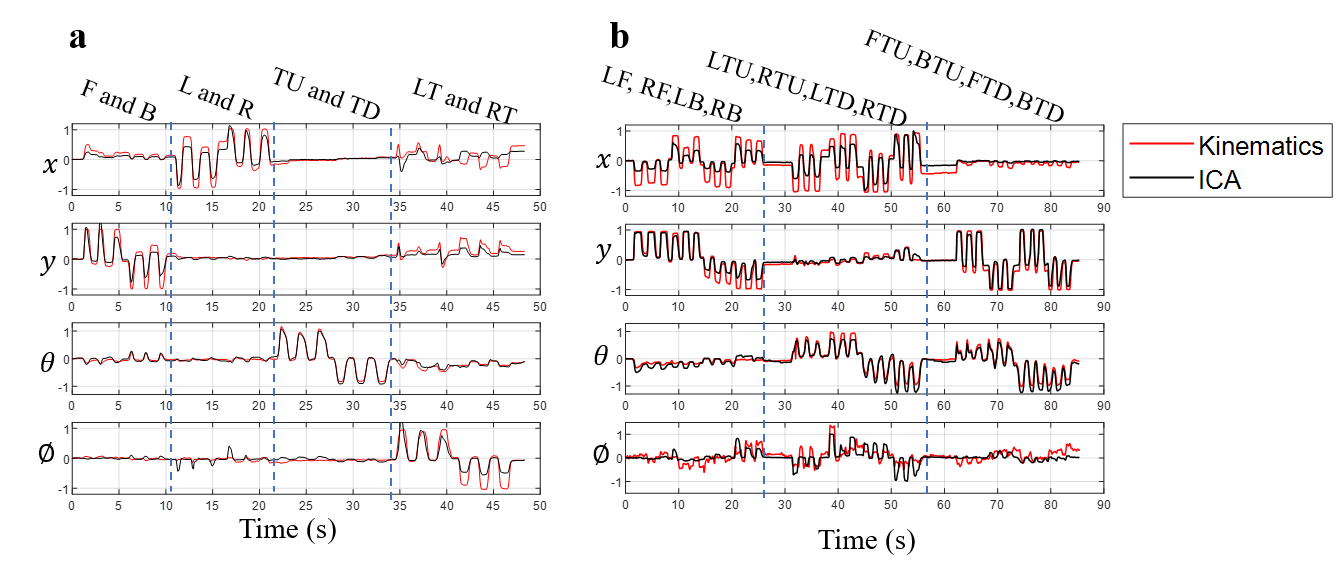}
\caption{Foot motion with kinematic transformation (red lines) and ICA (black lines) on testing data for subject 9 in single Cartesian directions data (a) and diagonal directions (b).}
\label{f:sub9}
\end{figure*}

Fig. \ref{f:sub9} further illustrates the results from the two mapping methods in four-DOF of testing data for subject 9. The sequences of motions for single and diagonal directions are labeled on the top of each figure. The ICA method effectively fixes the issue of inaccurate motions coupling in multi-DOF, for example, the directions of F and RT in Fig. \ref{f:sub9}a, and BTU in Fig. \ref{f:sub9}b.

To check the accuracy of control commands for all subjects and directions, the  {\it direction identification accuracy}, defined as the ratio of the sum of the number of movements applied in the desired direction to that the sum of the movements in all the directions, was used as a metric to test the performance of different mapping methods. For example, when forward direction is desired, correct sample data corresponding to \([x = 0 \, \wedge \, y > 0 \, \wedge \, \phi = 0 \, \wedge \, \theta = 0]\), while all motions correspond to the data satisfies \([x \neq 0 \, \vee \, y \neq 0 \, \vee \, \phi \neq 0 \, \vee \, \theta \neq 0]\).  

The diagonal directions can be regarded as rotated single directions and follow similar identification requirements in a transformed coordinate. For instance, when the target direction is LF, the data will be transformed with an anti-clockwise rotation of $45^o$ in $x-y$ plane, changing the target direction to negative axis of \(x\). The DOF of \(\phi\)  are not counted for diagonal direction analysis. As it cannot reach the extreme value during diagonal motions, the normalized data lose the relative relationship with the other DOFs. Thus, the correct sample data for direction LF should satisfy \([x < 0 \, \wedge \, y = 0 \, \wedge \, \theta = 0]\), and all motions data identified by \([x \neq 0 \, \vee \, y \neq 0 \, \vee \, \theta \neq 0]\).  

A defined zero band based on modeling data was set for the control commands in each DOF. The upper and lower limits in translations \(x,y\) and rotations \(\phi, \theta\) are identified as the 30\% and 40\% of the respective maximum and minimum values of the mapping results in kinematics and ICA. The zero band ranges are small: $[-0.6,0.6]cm$ for \(x,y\), $[-5^o,5^o]$ for $\phi$ and $ [-4^o,4^o]$ for \(\theta\) when reported in the theoretical kinematic model. For the subsequent testing data, if a mapping's result falls into the zero band, it is regarded as zero output in the responding DOF.
\begin{figure*}
\centering
\includegraphics[width=0.95\textwidth]{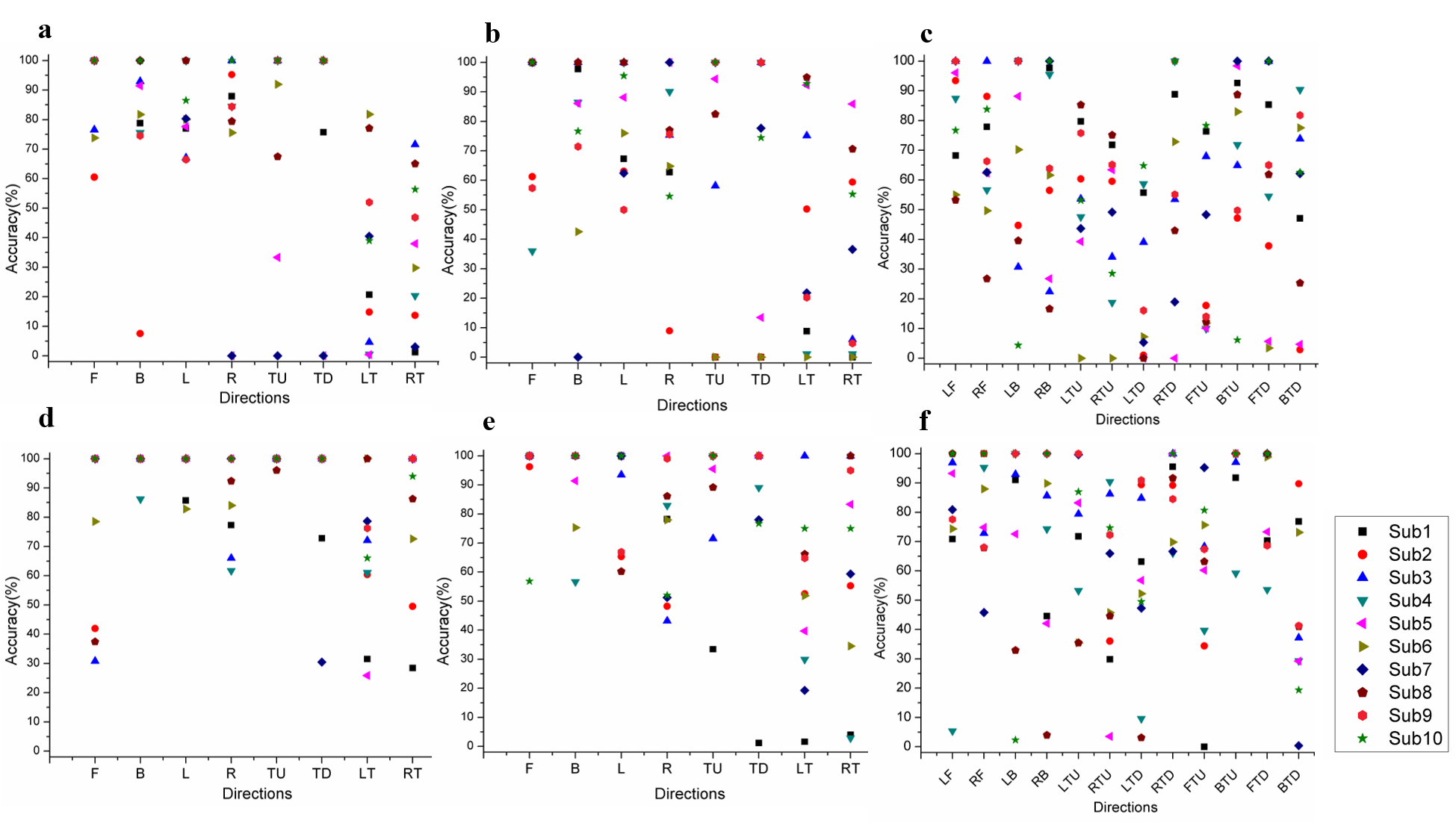}
\caption{Direction identification accuracy using the kinematic model transformation (a,b,c) and ICA model (d,e,f). (a,d) applies on the single Cartesian modeling data, (b,e) on the single Cartesian testing data, and (c,f) on the diagonal directions testing data.}
\label{f:ICAcompare} 
\end{figure*}
Fig. \ref{f:ICAcompare} shows the direction identification accuracy results of kinematic (Fig. \ref{f:ICAcompare}a-c) and ICA modeling (Fig. \ref{f:ICAcompare}d-f) for the 10 subjects of previous experiment. We see that the performance with the kinematic mapping is  improved using the subject-specific ICA mapping. However, even the ICA mapping cannot prevent specific problems such as subject 1 in the RT direction. Furthermore, the accuracy is lower in diagonal directions as expected. 

\section{Conclusion}
This paper introduced a four-DOF foot-controlled human-machine interface featuring force and position feedbacks, continuous output space control, and automatic home positioning. The design and modeling of this foot interface were presented, and an experimental study was conducted to quantify the performance of this interface. An approach was proposed to define a suitable mapping from foot movement space to output space based on ICA. 

The experimental data obtained with ten able-bodied subjects exhibited subject-specific movement patterns with obvious variability among different subjects but less variability in the repeated trials of each subject. This motivated us to develop a subject-specific mapping from the movement to an output command space using ICA, which  improved the control as compared with the common kinematic transformation approach. With the approach based on the ICA transformation, the accuracy of multiple directions over ten subjects increased (relatively to the kinematic transformation) from \(68\% \pm 16\%\) to \(88\% \pm 6\%\),  \(63\% \pm 19\%\) to  \(79\% \pm 12\%\), and  \(57\% \pm 13\%\) to \(72\% \pm 10\%\) for three datasets of single and diagonal directions respectively.

The above results on foot motion direction identification indicate that the foot interface system with built-in ICA model is able to identify multi-directions foot motion intentions accurately. 
Nevertheless, the performance of the ICA method largely depends on the calibration procedure and data, which should best reflect the habitual motion pattern of the specific subject. This is a possible reason why e.g. subject 1 cannot achieve a better result as the other subjects. 

\section*{Acknowledgment}
The authors would like to thank Jonathan Eden for his suggestions on improving this manuscript and the subjects who participated in this study. This work was funded by the Singapore National Research Foundation through the NRF Investigatorship
Award (NRF-NRFI 2016-07).

\bibliographystyle{IEEEtran}

\bibliography{IEEEabrv,Reference}
\begin{IEEEbiography}
[{\includegraphics[width=1in,height=1.25in,clip,keepaspectratio]{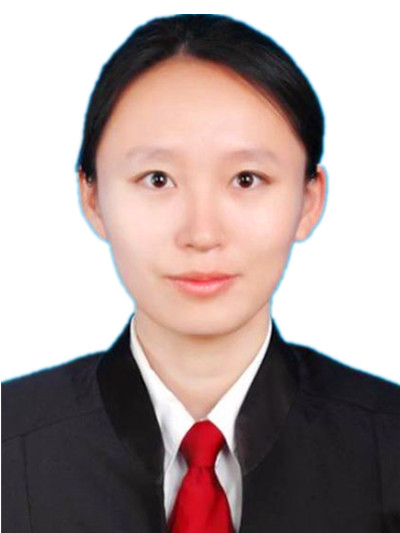}}]{Huang Yanpei} received her M.Sc. degree in Manufacturing Systems and Engineering from the School of Mechanical and Aerospace Engineering, Nanyang Technological University, Singapore, in 2016. She is currently pursuing her PhD study at the School of Mechanical and Aerospace Engineering, NTU in the field of human-machine interaction in robotic surgery.
\end{IEEEbiography}
\vskip 0pt plus -1fil
\begin{IEEEbiography}
[{\includegraphics[width=1in,height=1.25in,clip,keepaspectratio]{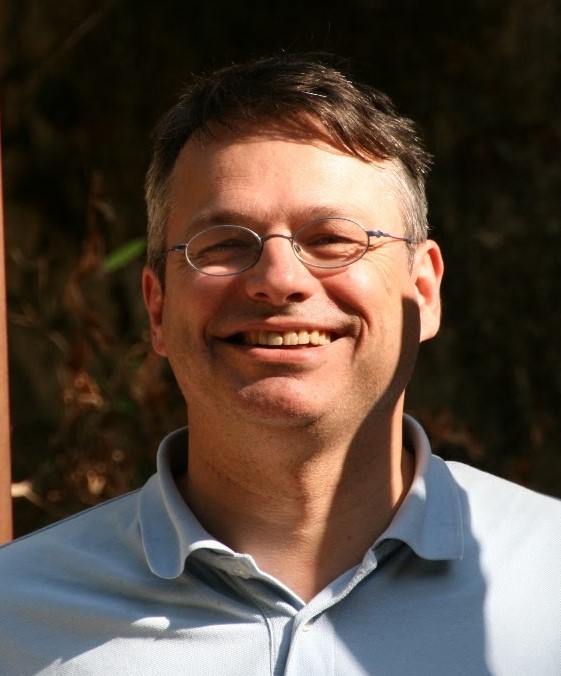}}]{Etienne Burdet}received the M.S. degree in mathematics, the M.S. degree in physics, and the Ph.D. degree in robotics, all from ETH-Zurich, Zurich, Switzerland, in 1990, 1991 and 1996, respectively. He currently is a Professor at Imperial College London, London, U.K. He is doing research at the interface of robotics and bioengineering, and his main research interest is human-machine interaction. He has contributions in various fields from human motor control to VR based training systems, assistive devices, and robotics for life sciences.
\end{IEEEbiography}
\vskip 0pt plus -1fil
\begin{IEEEbiography}
[{\includegraphics[width=1in,height=1.25in,clip,keepaspectratio]{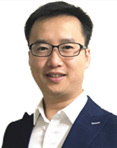}}]{Lin Cao} obtained his Ph.D. degree in Mechanical Engineering from University of Saskatchewan, Canada, in 2015. He is currently a Research Fellow in the Robotics Research Centre at Nanyang Technological University, Singapore. He was also a visiting scholar at the Department of Precision and Microsystems Engineering at Delft University of Technology, the Netherlands, in 2013. Dr. Cao's research interests include flexible endoscopic surgical robots, medical instruments, soft robots, haptics, and compliant mechanisms. He has developed the world's first robotic suturing system for flexible endoscopic surgery.
\end{IEEEbiography}
\vskip 0pt plus -1fil
\begin{IEEEbiography}
[{\includegraphics[width=1in,height=1.25in,clip,keepaspectratio]{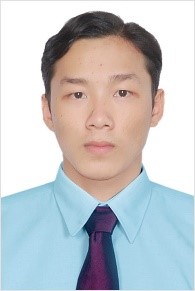}}]{Phuoc Thien Phan} received the B.Eng. degree in Mechanical Engineering from Vietnam National University, Ho Chi Minh city University of Technology (HCMUT) in 2015. He is currently a Research Assistant at School of Mechanical and Aerospace Engineering, Nanyang Technological University, Singapore. His research interests include the development of magnetic soft endoscopic capsule for obesity treatment, and mechanical design and development of novel medical devices, flexible surgical robotic systems.
\end{IEEEbiography}
\vskip 0pt plus -1fil
\begin{IEEEbiography}
[{\includegraphics[width=1in,height=1.25in,clip,keepaspectratio]{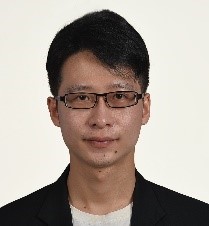}}]{Anthony Meng Huat Tiong} obtained his Master of Aeronautical Engineering (MEng) from Imperial College London, UK in 2015. He is currently a Research Associate in the Robotics Research Centre at Nanyang Technological University, Singapore. Mr. Anthony's research interests include deep learning, data analytics, and flexible endoscopic surgical robots. 
\end{IEEEbiography}

\vskip 0pt plus -1fil
\begin{IEEEbiography}
[{\includegraphics[width=1in,height=1.25in,clip,keepaspectratio]{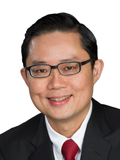}}]{Soo Jay Phee} graduated from Nanyang Technological University, Singapore (NTU) with the B.Eng (Hons) and M.Eng degrees in 1996 and 1999 respectively. He obtained his Ph.D. from Scuola Superiore Sant'Anna, Pisa, Italy in 2002 on a European Union scholarship.He is currently the Dean of the College of Engineering, NTU. Prior to his Deanship, he has served 3 years as Chair of the School of Mechanical \& Aerospace Engineering. Prof Phee's research interests include medical robotics and mechatronics in medicine. He is currently a recipient of the prestigious National Research Foundation (NRF) Investigator Award. Prof Phee had previously served as the Program Manager of A*STAR's inaugural MedTech Program. He was the founding CEO of EndoMaster Pte Ltd, a company he co-founded to commercialize a surgical robotic system he developed. 
\end{IEEEbiography}
\end{document}